\documentclass[conference]{IEEEtran}

\usepackage{cite}
\usepackage{amsmath,amssymb,amsfonts}
\usepackage{algorithmic}
\usepackage{graphicx}
\usepackage{textcomp}
\usepackage[usenames,dvipsnames]{color}
\usepackage{xcolor}

\def\BibTeX{{\rm B\kern-.05em{\sc i\kern-.025em b}\kern-.08em
    T\kern-.1667em\lower.7ex\hbox{E}\kern-.125emX}}

\makeatletter
\newcommand{\linebreakand}{%
  \end{@IEEEauthorhalign}
  \hfill\mbox{}\par
  \mbox{}\hfill\begin{@IEEEauthorhalign}
}
\makeatother    
\usepackage{bm}
\usepackage{stackrel}
\usepackage[switch]{lineno}
\begin{document}

\title{
Inverse Laplace Transform for Dynamic Light Scattering:
Impact of Regularization 
}

 \author{
 \IEEEauthorblockN{Pierre Pajuelo}
\IEEEauthorblockA{ENS de Lyon, CNRS, LPENSL \\
UMR5672, Lyon, France\\
pierre.pajuelo@ens-lyon.fr}
\and
\IEEEauthorblockN{Stéphane Roux}
\IEEEauthorblockA{ENS de Lyon, CNRS, LPENSL \\
UMR5672, Lyon, France\\
stephane.roux@ens-lyon.fr}
\and
\IEEEauthorblockN{Adrien Meynard}
\IEEEauthorblockA{ENS de Lyon, CNRS, LPENSL \\
UMR5672, Lyon, France\\
adrien.meynard@ens-lyon.fr}
\linebreakand
\IEEEauthorblockN{François Liénard}
\IEEEauthorblockA{ESRF—The European Synchrotron \\ Grenoble, France \\ francois.lienard@esrf.fr}
\and
\IEEEauthorblockN{Éric Freyssingeas}
\IEEEauthorblockA{ENS de Lyon, CNRS, LPENSL \\
UMR5672, Lyon, France\\
eric.freyssingeas@ens-lyon.fr}
\and
\IEEEauthorblockN{Pierre Borgnat}
\IEEEauthorblockA{CNRS, ENS de Lyon, LPENSL \\
UMR5672, Lyon, France\\
pierre.borgnat@ens-lyon.fr}
}

\maketitle

\begin{abstract}

Dynamic Light Scattering (DLS) analyzes particle dynamics from the autocorrelation functions of scattered light intensity, yet extracting accurate relaxation time distributions from noisy data is challenging. We develop an inverse problem approach to recover this distribution by inverting the Laplace transform with physics-based regularization, called the CONTIN method. We improve it to use it on noisy data, across a wide range of time scales, with a selection of the regularization strength through a data-driven L-curve criterion. Our approach enhances robustness under high noise and reveals multi-scale dynamics in complex systems. Validation is performed on simulated data, compared to the Cramér-Rao bound and to parametric methods, and on experimental data from Carbopol microgels. It demonstrates superior accuracy over parametric methods, especially for broad time distributions. The algorithm’s logarithmic discretization and variance-reduced correlation estimation enhance performance, offering a powerful tool for non-parametric DLS analysis and deeper insights into soft matter dynamics.

\end{abstract}

\begin{IEEEkeywords}
Light scattering, Inverse problem, Laplace transform, Non-parametric analysis.
\end{IEEEkeywords}

\vspace{-.15cm}
\section{Introduction}
\vspace{-.1cm}
Dynamic light scattering (DLS) is a non-invasive and non-destructive spectroscopic analysis technique for studying the dynamics of dispersed systems. It is primarily used to measure submicron particle sizes ($1\,\mathrm{nm}$ - $1\,\mu\mathrm{m}$). For over fifty years, DLS has been a standard tool in soft matter physics for extracting characteristic times and probing the dynamic properties of various soft matter systems \cite{BernePecora}.
%
DLS works by shining a light beam through a sample and measuring the scattered intensity at a given angle as a function of time, $I(t)$. For systems at thermodynamic equilibrium, i.e. ergodic, temporal correlations of $I(t)$ provide direct access to system dynamics \cite{BernePecora}. 

In practice, however, the measured signal is noisy, exhibits wide characteristic time distributions, is non-Gaussian, and possibly non-stationary for out-of-equilibrium systems. 
Recovering the distribution of relaxation times from the autocorrelation is therefore a non-trivial inverse problem. Existing approaches include parametric methods \cite{Frisken}, non-parametric inversions \cite{PROVENCHER1982229,Nieves}, or Bayesian formulations \cite{boualemthesis,boualemeusipco}. Yet, they do not provide a sufficiently accurate representation of the correlation functions in such challenging contexts.

We propose a modified version of the non-parametric CONTIN method, providing a way to specify the choice of regularisation and allowing us to accurately access the characteristic time content of correlation functions.
Section \ref{sec:2} introduces the physical model, the analysis with the Laplace transform, and the parametric and non-parametric methods. Our contributions in Section \ref{sec:3} are an improved estimator of the autocorrelation function, a numerical implementation adapted to a large scale of time lags (or delays), and a data-driven optimization approach to select the regularization strength. Section \ref{sec:4} shows numerical results on simulations and on experimental data, compared to parametric estimates.

\vspace{-.1cm}
\section{Background}
\label{sec:2}
\vspace{-.1cm}
\subsection{Autocorrelation Functions of DLS}
For DLS, two autocorrelation functions (ACF) are relevant: 
\begin{enumerate}
\item the normalised ACF of the electric field $E(t)$, denoted $g^{(1)}$, defined as
\begin{equation}
g^{(1)}(\tau) = \frac{\langle E^{\star}(t)E(t+\tau)\rangle }{\langle E^{\star} (t)E(t)\rangle },
\end{equation}
where $\tau$ is the delay and $\langle \cdot \rangle $ is a temporal average\footnote{In the case of a system in thermodynamic equilibrium, which we assume to be ergodic, this quantity is strictly equal to an ensemble average.};
\item the normalised ACF of the intensity $I(t)$, denoted $g^{(2)}$, defined as
\begin{equation}
g^{(2)}(\tau) = \frac{\langle I(t) I(t+\tau)\rangle}{\langle I(t)\rangle^2},
\end{equation}
where $I(t)=E^{\star} (t)E(t)$ is the intensity scattered in a direction defined by the scattering angle $\theta$.
\end{enumerate}

In a homodyne experiment, we get $g^{(2)}$. Indeed, the avalanche diode only provides access to the intensity, which constitutes a major difference compared to a heterodyne setup, which allows access to the function $g^{(1)}$ by interference.


Information regarding the system’s dynamics and its physical models is provided by the normalised autocorrelation function of the electric field, $g^{(1)}$. Under the assumptions that the total electric field follows Gaussian statistics, the number of scattering objects $N$ is large, and the light source is spatially coherent, chaotic and polarised, the two ACFs are related by
\begin{equation}
    g^{(2)}(\tau) = 1+\beta \,\lvert g^{(1)}(\tau)\rvert ^2.
    \label{eq:Siegert}
\end{equation}
This relation, known as the Siegert relation \cite{Siegert}, involves the parameter $\beta$, which accounts for the reduction of contrast when the measured intensity is averaged over a set of modes, or uncorrelated speckles. In the case where only a single spatial mode is collected and a single polarisation is selected, we expect to have $\beta=1$. Thus, thanks to this relation, we can theoretically trace back to the electric field autocorrelation function from the intensity autocorrelation measurement.

\subsection{Inverse Laplace Transform (ILT)}

In DLS, intensity is governed by fluctuations of the optical index, $n$. For a system composed of particles in solution with no interactions, and for a single relaxation mode, one can show that $g^{(1)}(\tau) \propto \exp(-q^2 D \tau)$, with $D$ the diffusion coefficient (of a single particle) and $q=\frac{4\pi n}{\lambda }\sin\left(\frac{\theta}{2}\right)$ the wave vector at angle $\theta$ and wavelength $\lambda$ \cite{BernePecora}. When several physical processes occur within a system, the ACF becomes a sum of exponential modes. 

This leads to a continuous approach where we model $g^{(1)}$ using a distribution $G(\Gamma)$ of relaxation rates $\Gamma$. Let $\mathcal{L}\{G\}$ be the Laplace transform of $G$. The continuous model reads:
\begin{equation}
    g^{(1)}(\tau) = \mathcal{L}\{G\}(\tau) :=\int_0^{+\infty} G(\Gamma)\exp(-\Gamma\tau)\,\mathrm{d}\Gamma, \label{eq:Laplacetransform}
\end{equation}
with $\int_0^{+\infty}G(\Gamma)\,\mathrm{d}\Gamma =1$ and $G(\Gamma)\geq 0$ for all $\Gamma\geq0$. Retrieving the characteristic time distribution $G$ thus requires performing the inverse Laplace transform (ILT) of $g^{(1)}$, a numerically ill-posed problem that the algorithm presented in section \ref{ssec:contin} is designed to address.

\vspace*{-.2cm}
\subsection{Parametric Estimate}
\vspace*{-.1cm}
Instead of estimating the distribution of relaxation rates $G(\Gamma)$, one can decide to perform a parametric approach, where we estimate the following quantities:
\begin{align}
    \bar{\Gamma} &= \int_0^{+\infty} G(\Gamma)\,\Gamma\,\mathrm{d}\Gamma,\\
    \mu_m = &\int_0^{+\infty}G(\Gamma)(\Gamma-\bar{\Gamma})^m\,\mathrm{d}\Gamma,
\end{align}
where $\bar{\Gamma}$ is the mean of $G$ and $\mu_m$ is the central moment of order $m$. By approximating the distribution $G(\Gamma)$ to a series expansion \cite{Koppel, HASSAN2006744, Mailer_2015}, and by injecting in eq. (\ref{eq:Laplacetransform}), one can deduce the series expansion of the ACF of $I$, using eq. (\ref{eq:Siegert}), as follows:

. 
\begin{equation}
\label{eq:g2parametric}
    g^{(2)} (\tau) = B+\beta \Big(\exp(-\bar{\Gamma}\tau)\big(1+\frac{\mu_2 \tau^2}{2}-\frac{\mu_3 \tau^3}{3!}+\dots)\Big)^2,
\end{equation}
where $B$ is the baseline, $\mu_2$ is the variance of the distribution $G(\Gamma)$, $\mu_3$ is related to the skewness coefficient
\cite{Frisken,Mailer_2015}. This formula allows us to estimate (and truncate) the moments of $G$ from observed data, by fitting a least-squares estimate of the parameters in eq.~(\ref{eq:g2parametric}).

\subsection{Solution by Inverse Problem}
\label{ssec:contin}
The CONTIN method is a generalised version of the algorithm initially proposed by S.W. Provencher \cite{Provencher79} and subsequently improved \cite{PROVENCHER1982213,PROVENCHER1982229}. The general purpose of the method is to solve linear integral equations of the first kind. 
Let us write $\| \cdot \|$ for the $\mathrm{L}^2$ norm, 
and $M$ the number of sampled time lags.  CONTIN provides an estimate $\hat{G}$ of the distribution $G$, as the solution to an inverse problem, formulated as the sum of a data fidelity term and a regularisation term, as follows: 
\begin{equation}
\hat{G} = \arg\min_{G\geq 0}\ \sum_{m=1}^{M} 
\left\| g^{(1)}(\tau_m) - \mathcal{L}\{G\}(\tau_m) \right\|^2 + \alpha \| R\cdot G \|^2,
\label{eq:ILTregul}
\end{equation}
where $\alpha$ is the regularisation parameter, or strength, and $R$ is the regularisation operator. 
Tikhonov regularisation would assume $R=\mathbb{I}$, the identity operator. However, we physically expect smooth solutions, coherent with the bell-shaped distributions, e.g. Gaussian, obtained when modeled by the parametric method. To favour such a solution, we choose for regularisation $R=\mathbb{D}^{(2)}$, the second derivative operator with respect to the variable $\Gamma$, and this better follows what is expected.

\vspace{-.2cm}
\section{Methodological Contributions}
\label{sec:3}
\vspace*{-.1cm}
\subsection{Estimation of $g^{(2)}$, ACF of Light Intensity}
\vspace*{-.1cm}
To estimate the distribution $G(\Gamma)$, we need access to a very wide range of delays, sufficiently sampled, and this is crucial for the numerical discretization of the integral associated with the Laplace transform. 
We must also find a correlation estimator that is robust to noise over a large range of lag time, in order to improve the performance of the estimation.  

Experimentally, only a finite observation time $T$ is available, due to memory and acquisition time constraints. As a result, correlation estimators must remain reliable for finite-length signals, either because of relatively short recording durations or possibly due to non-stationary components that preclude the use of long windows, and with delays going up to $T$. Direct time-domain estimation of the correlation function by summing all time-shifted pairs is computationally expensive. 
It is useful to switch to the frequency domain, which offers a double advantage: reduced computation load for the convolution thanks to the Fast Fourier Transform (\texttt{fft}) algorithm, reduced variance by constructing such an estimator for the spectrum. 
Let $S$ denote the power spectral density of the signal $I(t)$, given by $S(\nu) = \frac1{T}\lvert \mathcal{F}\{ I\}(\nu)\rvert^2$ where $\mathcal{F}$ denotes the temporal Fourier transform. According to the Wiener–Khinchin theorem for stationary processes:
\begin{equation}
    g^{(2)}(\tau) = \dfrac{\mathcal{F}^{-1}\{S\}(\tau)}{\langle I(t) \rangle^2},
\end{equation}
Computing $g^{(2)}$ in this way leverages the \texttt{fft} numerically, and reduces computation time to an order $\mathcal{O}(M \log M)$.

To obtain a more robust estimator of $g^{(2)}$, we estimate the power spectrum $S$ using Welch's periodogram. 
%
%
Apodising the observed signal $I(t)$ with a window $\omega_T(t)$ yields an asymptotically unbiased estimator as the observation time $T$ tends to infinity \cite{papoulis_signal_1977}. Nevertheless, its variance scales as $\lvert S(\nu)\rvert^2$, and does not decrease with increasing $T$.
Welch’s method reduces this variance by dividing the signal into $K$ overlapping segments of length $W\ll T$, computing individual spectra, and averaging them:
\begin{equation}
    \hat{S}_{\mathrm{p}}(\nu) = \frac{1}{KW}\sum_{k=0}^{K-1} \lvert \mathcal{F}\{y_k\}(\nu) \rvert^2,
\end{equation}
where $y_k(t) = I(t-kW)\,\omega_W(t)$. This procedure increases the bias because the support of $y_k(t)$ is now of duration $W$ smaller than that of the full signal $I(t)$. The variance of this estimator behaves as $\lvert S(\nu)\rvert^2/K$, providing a substantial variance reduction for large $K$ \cite{papoulis_signal_1977}.

\vspace{-.3cm}
\subsection{Numerical Implementation of ILT for Wide Range of Times}

In the original CONTIN method, the minimisation problem \eqref{eq:ILTregul} is linearly discretised, because the delays $\tau$ and the characteristic times $\Gamma^{-1}$ are supposed to be in a small  range of possibilities.  
For complex fluids, we cover a larger range and we discretize \eqref{eq:ILTregul} over a logarithmically spaced  grid $\{\Gamma_i\}_i$. The objective is to estimate $\bm{G}$, the discrete version of the relaxation rate distribution $G$. We approximate the Laplace integral using a trapezoidal scheme, yielding
\begin{equation}
\mathcal{L}_{\mathrm{trap.}}\{\bm{G}\}(\tau_i) = \sum_j \tilde{A}_{ij}G_j \frac{\log_{10}\left(\Gamma_{j+1}\right)-\log_{10}\left(\Gamma_{j-1}\right)}{2},    
\end{equation}
where $\tilde{A}_{ij} = \exp\left(-\tau_i \Gamma_j \right)$. Let us introduce the vector $\bm{b}$ such that $b_i = \lvert g^{(1)}(\tau_i) \rvert^2$. Thanks to eq.~(\ref{eq:Siegert}), $b_i$ equals the normalised and centered ACF of $I(t)$. Defining the matrix $\bm{A}$ whose elements are given by $A_{ij}=\tilde{A}_{ij} \left(\log_{10}\left(\Gamma_{j+1}\right)-\log_{10}\left(\Gamma_{j-1}\right)\right)/2$, the regularised inverse problem becomes
\begin{align}
    \bm{\hat{G}}_2 &= \arg \stackrel[G_i \geq 0]{}{\min} \Big(\| \bm{A} \bm{G} - \bm{b} \|^2 + \alpha \| \bm{\mathbb{D}}^{(2)}_{\mathrm{num}}  \bm{G}\|^2\Big), \\
    & =  \arg \stackrel[G_i \geq 0]{}{\min} \Big(\| \bm{C} \bm{G} - \bm{d} \|^2 \Big), \bm{C} = \begin{bmatrix}
      \bm{A} \\ \alpha \,\bm{\mathbb{D}}^{(2)}_{\mathrm{num}}  \bm{G}
    \end{bmatrix}, \bm{d} = \begin{bmatrix}
      \bm{b} \\ \bm{0}
    \end{bmatrix}
    \label{eq:ILTalgo}
\end{align}
with $\bm{\mathbb{D}}^{(2)}_{\mathrm{num}} = (\mathbb{I}_{-1}+\mathbb{I}_{+1}-2\,\mathbb{I})/ (\log_{10}(\Gamma_{+1})-\log_{10}(\Gamma))^2$, and $X_{+1 (-1)}$ is the vector $X$ shifted by one step forward (backward resp.). 
In this formulation, we indicate by the index ``2'' in $\hat{\bm{G}}_2$ that the inverse problem is on $\lvert g^{(1)} \rvert^2$ (and related to $g^{(2)}$). The initial problem would have a solution $\hat{\bm{G}}_1$ by doing the ILT of $g^{(1)}$. The two distributions are linked as, in practice, we have $G_1(\Gamma)  \propto G_2(2 \Gamma)$ -- this is exact for a single mode and discussed in \cite{lienard_multiscale_2022} for more modes.

The distribution is finally converted back to linear space according to $\bm{G}_{\mathrm{lin}}(\Gamma)$ as $\bm{G}_{\mathrm{lin}}(\Gamma) = \bm{G}(\log_{10} \Gamma)
\mathrm{d}\log_{10}\bm{\Gamma}/\mathrm{d}\bm{\Gamma}$, resulting from the change of variables $\gamma = \log_{10} \Gamma$.

For the numerical implementation, we rely on the non-negative least square algorithm \cite {lawson_solving_1995, Afrough}, implement in Python in the SciPy library as \texttt{scipy.optimize.nnls}. 

The Hanning window is used for the Welch periodogram, and the window number $K=10$ is chosen such that the maximum delay available on the ACF is not drastically reduced (a factor 10), with sufficient reduction of the variance. 
\vspace{-.3cm}
\subsection{Data-Driven Choice of Regularisation Parameter}
While the choice of the regularisation strength $\alpha$ may appear arbitrary, we rely here on an approach that optimises it from a data-driven criteria. We briefly recall here the L-Curve method \cite{HansenLCurve} that we implement to determine the optimal value, for a regularisation operator $R=\mathbb{D}^{(2)}$. 

\begin{figure}[htbp]
    \centering
    \includegraphics[width=1\linewidth]{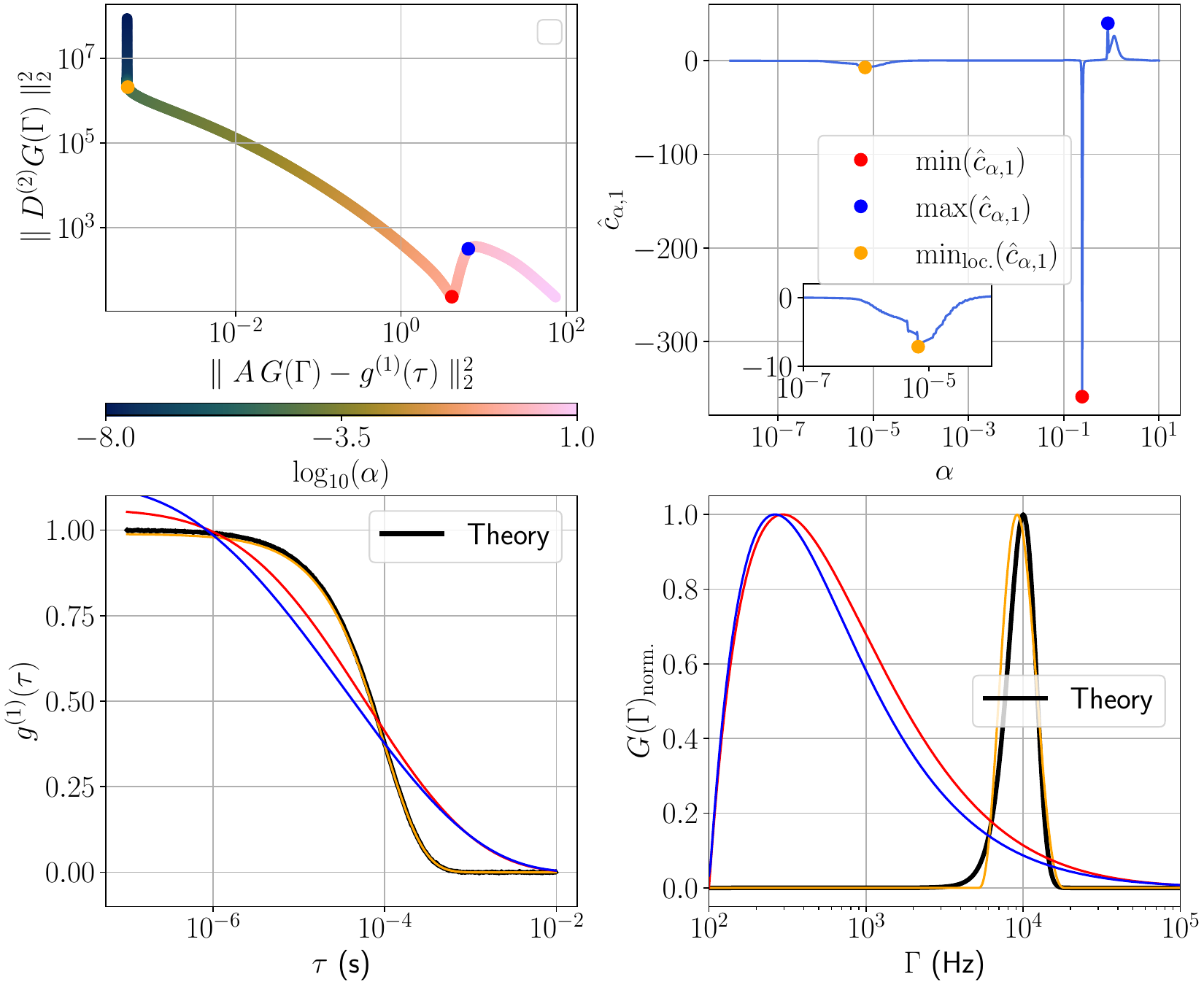}
    \caption{\textbf{(Top left)} L-Curve; \textbf{(Top right)} Estimated curvature of L-Curve; \textbf{(Bottom left)} ACF of $E$ as function of time lag, in black the curve simulated with $\sigma_b = 10^{-3}$, in red, blue and orange the curves obtained by our method, for the different extremum in curvatures; \textbf{(Bottom right)} Normalised relaxation distribution $G_{\mathrm{norm.}} = G_1(\Gamma)/\mathrm{max}_{\Gamma}G_1(\Gamma)$ as function of relaxation rate; in black the distribution simulated with $\bar{\Gamma} = 10^4, \sigma =2\times 10^3$, in red, blue and orange the estimated distributions; the best fit is for the orange point.}
    \label{fig:CONTINworkline}
\end{figure}

The principle is to obtain a trade-off between the data fidelity term and the regularisation term \cite{HansenLCurve}.
By plotting the two terms of the cost function \eqref{eq:ILTregul} (one as a function of the other), for several values of $\alpha$ (spaced logarithmically), in log-log representation, exhibits an ‘L’-shaped curve, hence its name. 
Denoting the regularisation term by $\xi$ and the data fidelity term by $\rho$, we define the curvature of the L-curve as
\begin{equation}
    \hat{c}_{\alpha} = 2\,\frac{\hat{\rho}' \hat{\xi}'' - \hat{\rho}''\hat{\xi}'}{\big((\hat{\rho}')^2 + (\hat{\xi}^2)\big)^{3/2}},
\end{equation}
where $\hat{\xi} = \log(\xi)$, $\hat{\rho} = \log(\rho)$ and primes denote derivatives with respect to $\alpha$. The curvature thus has extrema (local or global), from which the optimal regularisation parameter can be identified. In practice, the procedure involves computing the inverse solution for a range of $\alpha$ values, constructing the corresponding L-Curve, and selecting $\alpha$ at one of the points of maximum curvature. We keep the maximum curvature point with smallest data fidelity norm. This is illustrated in Fig. \ref{fig:CONTINworkline}. 

\vspace*{-.4cm}
\section{Validation}
\label{sec:4}

\subsection{Simulated Data}
We generate synthetic first-order correlation functions, at discrete lag times $\{\tau_m\}_{m=1}^M$, spaced logarithmically. The noisy observations, $\tilde{g}^{(1)}(\tau_m)$, are modelled as
\begin{equation}
    \tilde{g}^{(1)}(\tau_m)  = g^{(1)}(\tau_m,\bm{\Theta}) + n(\tau_m)
    \label{eq:noisy.model}
\end{equation}
where $g^{(1)}(\tau_m,\bm{\Theta})$ is the noise-free correlation function, defined as the Laplace transform (see eq. \eqref{eq:Laplacetransform}) of a single Gaussian mode parameterised by its mean $\bar{\Gamma}$ and variance $\sigma$, so that $\bm{\Theta} = [\bar{\Gamma}, \sigma]^{\mathrm{T}}$. Here, $n(\tau_m)$ is an additive Gaussian white noise, with zero mean and variance $\sigma_{n}$.

To assess the statistical performance of the estimators, we recall the Cramér-Rao bounds (CRB) for estimates of the parameters of model \eqref{eq:noisy.model}, grouped in the vector $\bm{\tilde{\Theta}} = [\bm{\Theta}, \sigma_n]^\mathrm{T}$. The CRBs are given by the diagonal elements of the inverse of the Fisher information matrix (FIM) \cite{boualemgretsi,boualemthesis}. The elements of the FIM are defined by 
\begin{equation}
    \mathcal{I}_{i,j} = - \mathbb{E}\left[\frac{\partial^2 \ell (\tilde{g}^{(1)}; \bm{\Tilde{\Theta}})}{\partial\Tilde\Theta_i\partial\Tilde\Theta_j}\right],\
\end{equation}
where $\ell (\tilde{g}^{(1)};\bm{\tilde{\Theta}})$ is the log-likelihood function, given by
\begin{align}
    \begin{split}
        \ell(\tilde{g}^{(1)};\bm{\tilde{\Theta}}) =&-\frac{M}{2}\log\left(2\pi \sigma_{n}^2\right)\\
        &-\frac{1}{2\sigma_{n}^2}\sum_{m=1}^{M} \left( \tilde{g}^{(1)}(\tau_m)-g^{(1)}(\tau_m, \bm{{\Theta}})\right)^2.
    \end{split}
\end{align}
Under the noise assumptions (zero mean, variance $\sigma_{n}^2$), the elements of the FIM are:
\begin{align}
\begin{split}
    &\mathcal{I}_{\Theta_i,\Theta_j} = \frac{1}{\sigma_{n}^2} \sum_{m=1}^{M} \frac{\partial g^{(1)}(\tau_m, \bm{{\Theta}})}{\partial \Theta_i}\frac{\partial g^{(1)}(\tau_m, \bm{{\Theta}})}{\partial \Theta_j},\\
    &\mathcal{I}_{\Theta_i,\sigma_{n}} = 0,\quad \mathcal{I}_{\sigma_{n},\sigma_{n}} = \frac{M}{2\sigma_{n}^4}.
\end{split}
\end{align}
The CRB for parameter $\Theta_i$ is given by $\mathrm{CRB}(\Theta_i) =(\mathcal{I}^{-1})_{i,i}$. For any unbiased estimator $\hat{\Theta}_i$ of $\Theta_i$, the mean square error (MSE) satisfies: 
\begin{equation}
\mathrm{MSE}(\hat{\Theta}_i)=\mathbb{E}\left[(\hat{\Theta}_i-\Theta_i)^2\right]\geq \mathrm{CRB}(\Theta_i).
\end{equation}
Since analytical inversion of the FIM is not tractable, the CRB is evaluated numerically,
following \cite{boualemgretsi,boualemthesis}.

To statistically evaluate the robustness of the proposed method, the estimation performance of the mean of the relaxation rate distribution is assessed using simulated data over the range $\tau_{\min} = 10^{-7}$ to $\tau_{\max}=10^{-2}$ spaced logarithmically with $n_{v,\tau}=100$ points per decade. The noise standard deviation $\sigma_n$ is varied from $10^{-4}$ to $0.5$. For each value of $\sigma_n$, we generate 500 Monte Carlo realisations of the noise.

\begin{figure}[t]
    \centering
    \includegraphics[width=1\linewidth]{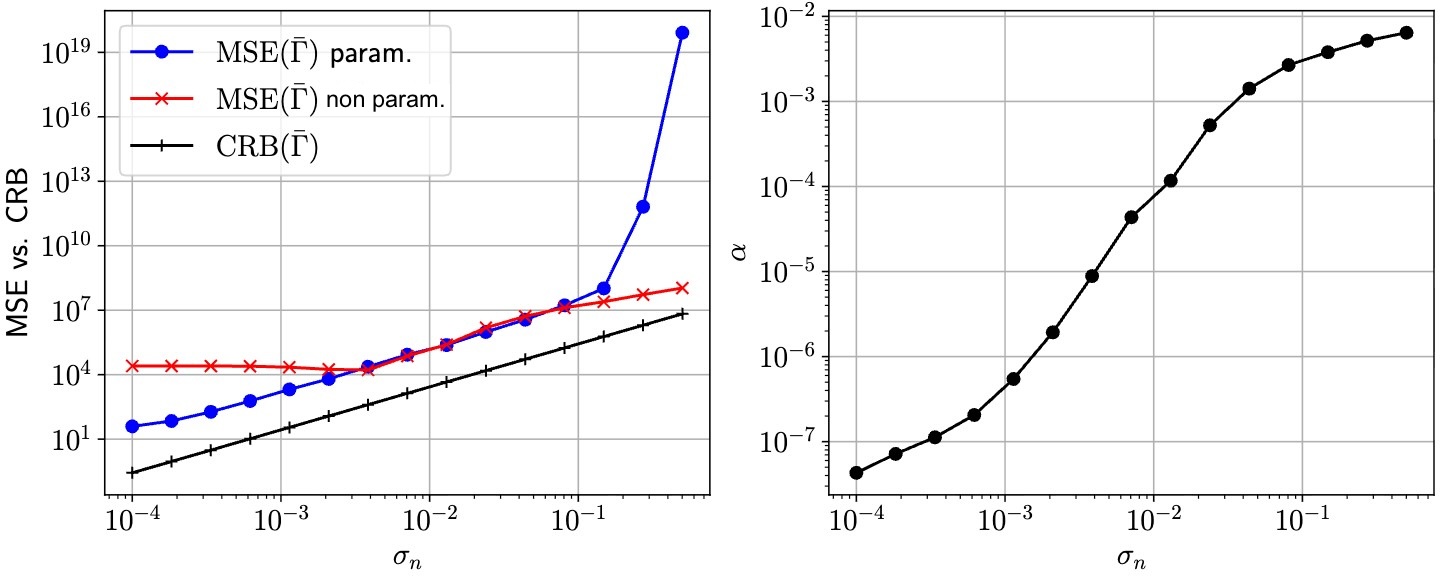}
    \caption{\textbf{(Left)} MSE and CRB as function of the standard deviation of the noise $\sigma_{n}$ for the averages of 500 realizations of the noise at each $\sigma_n$; parameters are $\bar{\Gamma}=10^4$ and $\sigma = 10^3$; \textbf{(Right)} Regularisation parameter as function of $\sigma_{n}$, averaged on 500 simulations; parameters are $\bar{\Gamma}=10^4$ and $\sigma = 10^3$.}
    \label{fig:MSECRB}
\end{figure}

We compute the MSE of the estimation of $\hat{\bar{\Gamma}}$ for the parametric estimator and for the modified {CONTIN} approach, and we plot it on Fig. \ref{fig:MSECRB}-Left. For each simulation, we select the optimal value of the regularisation parameter via the L-curve curvature criterion, and its average value over the 500 realisations is reported on Fig. \ref{fig:MSECRB}-Right.

The results show that neither method reaches the CRB. The proposed method shows a plateau, insensitive to $\sigma_n$ at very small noise levels, with a larger MSE than the parametric approach. However, for sufficiently high noise levels, $\sigma_n \gtrsim 3\times 10^{-3}$, the modified CONTIN method becomes more robust than the parametric estimator regarding the estimation of $\bar{\Gamma}$. The optimal regularization parameter increases by approximately three orders of magnitude when the noise variance increases by two orders of magnitude, highlighting the critical role of regularisation in noisy regimes.

\subsection{Experimental Data on Carbopol Solution}

Carbopol, an acrylate-based family of cross-linked, high molecular weight synthetic polymers, is a thickener widely used in industrial products. This type of polymer swells in water to form a gel. Physical properties of these gels depend on Carbopol concentration and pH. The mechanical response of these systems are well known, but their microstructures and dynamics have yet to be studied \cite{LeeFrisken, Divoux}. 

\begin{figure}[t]
    \centering
    \includegraphics[width=1\linewidth]{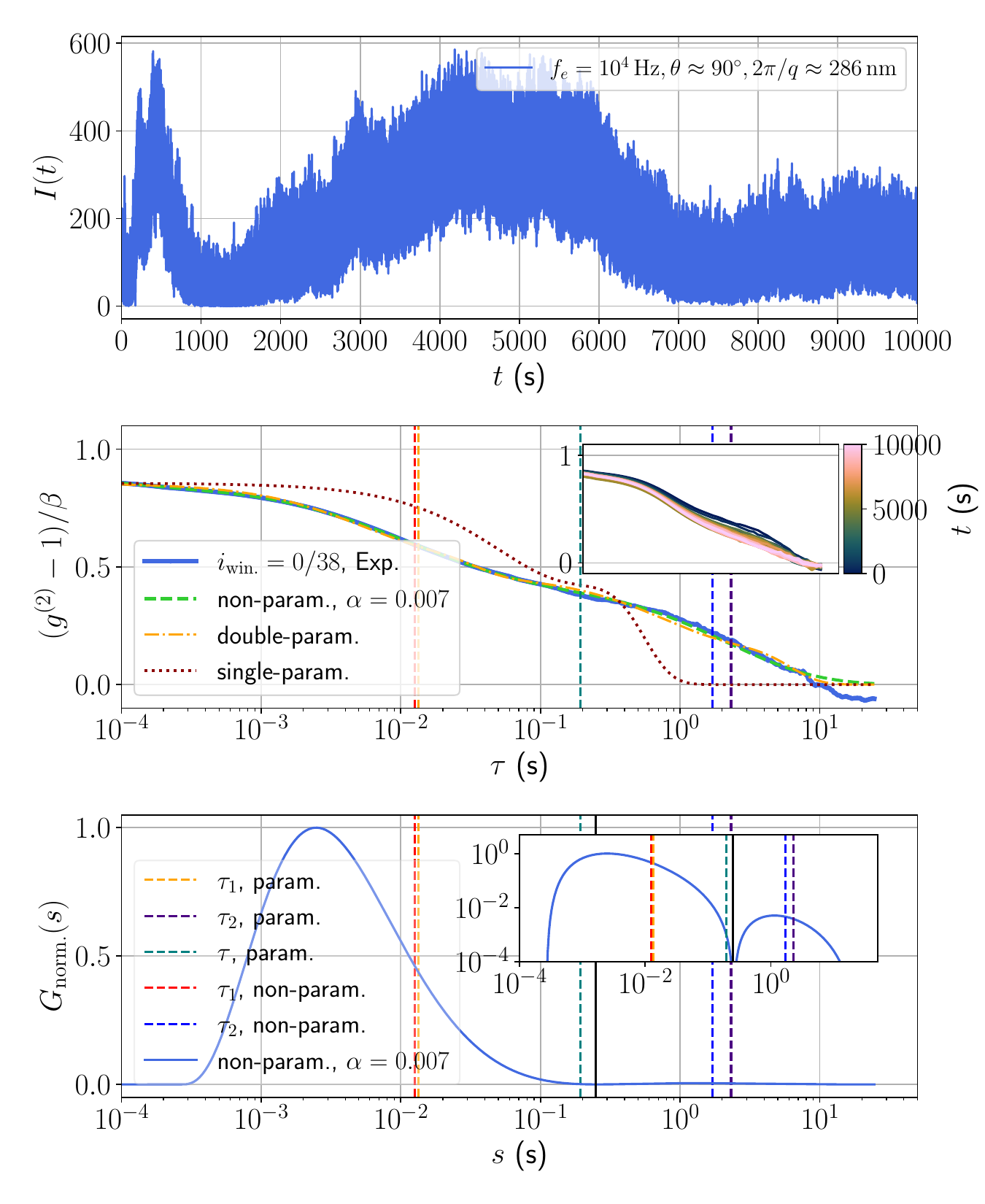}
    \vspace*{-0.7cm}
    \caption{(\textbf{Top}) Scattered intensity as function of time; (\textbf{Middle}) Normalised ACF of the as function of the lag time, in blue for the first window of size $W$ in the signal of size $T$, in dashed green for the fit from the proposed non-parametric method, in dashed brown and orange the fit from parametric estimate with one or two characteristic times ; (\textbf{Middle insert}) Normalised ACF of intensity as function of the lag time, for different windows of size $W=500$s (the color is the starting time $t$ of each window); (\textbf{Bottom}) Normalised relaxation distribution $G_{\mathrm{norm.}}(s)= \tilde{G}_2(s)/\max_{s}\tilde{G}_2(s)$ as function of relaxation time $s$, in coloured dashed lines the characteristic times extracted from the parametric and the proposed methods, in black line the split of the integrals associated with the calculation of the mean of each distribution  
    ; (\textbf{Bottom insert}) Log-log representation of the normalised relaxation distribution.}
    \label{fig:carbopol}
\end{figure}

To validate the proposed algorithm on experimental data, we set up a DLS device using a vertically polarised laser with wavelength $\lambda=532\,\mathrm{nm}$, and study the intensity fluctuations in a direction making an angle $\theta$ with the incident beam. We placed a sample of Carbopol EDT-2050 (weight concentration of 5\% and $\mathrm{pH}\sim 6$) into the DLS device at room temperature and recorded the scattering intensity over time (Fig. \ref{fig:carbopol}-Top panel). As this sample is a gel, it is non-ergodic. It is therefore a suitable system to test our method.

The acquisition duration is $T = 10\,000\,\mathrm{s}$, with a sampling frequency $f_s = 10^4 \,\mathrm{Hz}$, enabling access to a broad range of correlation times. We divide the signal into windows of duration $W=500\,\mathrm{s}$ with 50\% overlap, in particular to highlight any potential changes in the system over the measurement period. We consider the system to be in quasi-equilibrium over the period $W$ so that we can analyse the correlation function and extract the characteristic physical times.

To display the physical characteristic times (rather rates) contained in the ACF, we introduce the change of variables $\Gamma = \frac1{s}$, yielding
\begin{equation}
    g^{(1)}(\tau) = \int_0^{+\infty} \tilde{G}(s)\exp(-s\tau)\,\mathrm{d}s,
\end{equation}
with $\tilde{G}(s) = G(1/s)/s^2$ the normalized relaxation rate distribution. For convenience, we denote this normalized relaxation time distribution by $G(s)$. 

The advantage of the proposed algorithm is that it allows us to discriminate, without any prior assumption, between a model with one or more parameters. Indeed, the fit obtained for the non-parametric approach suggests two broad characteristic time distributions separated by several orders of magnitude. The corresponding fit yields $\mathrm{RMSE} = 0.172$, consistent with an effective noise level $\sigma_n \sim 0.2$. In this regime, the numerical validation (Fig. \ref{fig:MSECRB}-left) indicates that the proposed method provides a lower mean square error on the estimated mean relaxation rate than the parametric alternative.


\section{Conclusion}
\vspace*{-.1cm}
In this study, we developed and implemented a modified version of {CONTIN} 
in which the regularisation strength estimated by optimising the $L$-curve. The numerical implementation proved to be effective and practical, yielding low variance in the estimation of the characteristic times of the ACF. The use of logarithmic discretisation allows us to handle relaxation-time distributions spanning several orders of magnitude.

A first perspective would be to compare the approach proposed here 
with stretched-exponential modelling. While the latter often provides excellent empirical fits with fewer parameters, it reduces the direct physical interpretability of the extracted quantities.
A second perspective would be to develop a full time-Laplace approach in order to optimise the size of the analysis window $W$ and ensure a consistent estimate of the correlation functions in potentially non-stationary regimes.
A last improvement would be to increase the frequency resolution (while keeping the reduced variance) by replacing the Welch method by a multi-taper approach \cite{MT}.  


\end{document}